\documentclass[journal]{IEEEtran}
\usepackage{amsmath,amsfonts}
\usepackage{algorithmic}
\usepackage{algorithm}
\usepackage{array}
\usepackage[caption=false,font=normalsize,labelfont=sf,textfont=sf]{subfig}
\usepackage{textcomp}
\usepackage{stfloats}
\usepackage{url}
\usepackage{verbatim}
\usepackage{graphicx}
\usepackage{cite}
\usepackage{gensymb}
\begin{document}
\title{ARINC 429 Cyber-vulnerabilities and Voltage Data in a Hardware-in-the-Loop Simulator}

\author{Connor~Trask, Steve~Movit, Justace~Clutter, Rosene~Clark, Mark~Herrera, Kelly~Tran% <-this % stops a space
\thanks{C. Trask et al. are with the Institute for Defense Analyses, Alexandria, VA, 22305 USA email: ctrask@ida.org}}% <-this % stops a space

\maketitle

\begin{abstract}
ARINC 429 is a ubiquitous data bus for civil avionics, enabling reliable communication between devices from disparate manufacturers. However, ARINC 429 lacks any form of encryption or authentication, making it an inherently insecure communication protocol and rendering any connected avionics vulnerable to a range of attacks. We constructed a hardware-in-the-loop simulator with ARINC 429 buses, explored these vulnerabilities, and identified their potential to deny, degrade, or disrupt aircraft capabilities. We performed a denial-of-service attack against a multi-function display via a compromised ARINC 429 bus using commercially available tools, which succeeded in disabling important navigational aids. This proven attack on physical avionics illustrates the risk inherent in ARINC 429 and the need for the ability to detect these attacks. One potential mitigation is an intrusion detection system (IDS) trained on data collected from the electrical properties of the physical bus. Although previous research has demonstrated the feasibility of an IDS on an ARINC 429 bus, no IDS has been trained on data generated by avionics hardware. To facilitate this, we recorded voltage traces and message history generated by avionics and adversarial devices on the ARINC 429 bus. To the best of our knowledge, this is the first publicly available collection of hardware-generated ARINC 429 signal data.
\end{abstract}


\IEEEpeerreviewmaketitle

\section{Introduction}

% The very first letter is a 2 line initial drop letter followed by the rest of the first word in caps.
\IEEEPARstart{A}{RINC} 429 is a serial line communication standard specifically designed for use in avionics\cite{arinc_std}, enabling reliable communication between line-replaceable units (LRUs) from different manufacturers. \par
The standard is ubiquitous in commercial aviation, having been first introduced in the late 1970s by Aeronautical Radio Incorporated (ARINC), and as a result can be found on most active and retired commercial aircraft \cite{Fuchs2012TheEO}. Although more modern standards such as Avionics Full-Duplex Switched Ethernet (AFDX/ARINC 664 \cite{afdx}) are available, ARINC 429 will likely remain in service on older aircraft and continue to be used in select capacities on new aircraft. \par
When ARINC 429 was first designed and implemented, reliable transmission of messages was critical, necessitating a highly deterministic protocol with low response times. Security was not an area of emphasis in the creation of the standard, however, ARINC 429 has been updated 19 times since its initial publication, most recently in 2019\cite{arinc_std}. These updates to the standard have added additional data words, labels, and equipment identifiers as new avionics are invented and incorporated into aircraft, but they have not added security measures or fundamentally changed the nature of the system. As a result, ARINC 429 still has none of the security features found in modern secure communication protocols, such as message encryption or authentication. Despite this lack of security, much of the existing literature on avionics cybersecurity either overlooks ARINC 429 or assumes it is secure due to its architecture.
ARINC 429 is a unidirectional communication protocol, where one transmitter LRU sends messages to no more than 20 receiver LRUs. If a receiver LRU wishes to send messages back to the transmitter LRU, this must be done over a physically separate ARINC connection, where the original receiver acts as the new transmitter. ARINC 429 messages consist of 32 bits sent in a fixed format using a bipolar return-to-zero encoding; ARINC 429's electrical characteristics\cite{arinc_std} are detailed in Table \ref{table:I}. The standard-specified cabling for ARINC 429 is a shielded twisted-pair, carrying a differential signal along the two wires that is summed at the receiver to eliminate noise induced on both wires.

\begin{table}[h!]
\centering
\caption{ARINC 429 Electrical Signal Characteristics \cite{arinc_std}}
\label{table:I}
\renewcommand{\arraystretch}{1.25}
\begin{tabular}{||c c||} 
 \hline 
 Binary State & Voltage Differential (A-B)  \\ 
 \hline\hline 
 "HI" - 1 & +6.5V to +13V \\ 
 "NULL" & +2.5V to -2.5V \\
 "LO" - 0 & -6.5V to -13V \\ 
 \hline
\end{tabular}
\end{table}

ARINC 429 messages are sent as 32-bit-long words. The first 8 bits are the label, followed by two bits for the source/destination identifier (SDI), 19 bits for the data field, two bits for the sign/status matrix, and a parity bit (P). The format of ARINC 429 messages is often presented in the way that it would appear when read on an oscilloscope, with the first bit received on the far right and the last bit received on the far left, as illustrated in Table \ref{table:II}

\begin{table*}[ht]
\centering
\caption{ARINC 429 Message Format \cite{arinc_std}}
\label{table:II}
\begin{tabular}{|c|c|c|c|c|c|c|c|c|c|c|c|c|c|c|} 
 \hline 
 32 & 31 & 30 & 29 & 28 & ... & 12 & 11 & 10 & 9 & 8 & 7 & ... & 2 & 1 \\ 
 \hline
 P & \multicolumn{2}{|c|}{SSM} & \multicolumn{1}{|c}{MSB} & \multicolumn{3}{c}{Data} & LSM & \multicolumn{2}{|c|}{SDI} & \multicolumn{5}{|c|}{Label}\\
 \hline
\end{tabular}
\end{table*}

The label is expressed in octal notation, with possible values from 000 to 377. The first bit of the label is the most significant one. The two-bit SDI field can be used to indicate the message's intended receiver, or to identify which subsystem is using the transmitter. The SDI field can also be used to add an extra two bits to the data field. The 19-bit data field is transmitted least significant bit first, following the Universal Asynchronous Receiver/Transmitter (UART) protocol \cite{uart}. The data format for an ARINC 429 message can consist of bit field discrete data, binary coded decimal (BCD), binary number representation (BNR), or a combination of the above. The two bits after the data field are the sign/status matrix, which indicates either the status of the transmitting device or the sign of the transmitted data, depending on the data format. Finally, ARINC 429 messages are odd parity, requiring that every valid message contains an odd number of high-level signals (ones). If the parity bit is set incorrectly, resulting in an even number of ones, this signals to the receivers that an error occurred with the transmitter or bus. The parity bit helps confirms the integrity of a transmitted message through simple error detection but does not provide enough information to correct bit errors in the message. ARINC messages can be transmitted at one of two fixed bit rates: slow (12 - 14.5 Kbits/s) or fast (100 Kbits/s); both rates are self-clocked. 

\subsection{Understanding ARINC 429 Security Vulnerabilities}
Compared to other avionics buses, there is little published knowledge on ARINC 429's security vulnerabilities. There is a large and robust body of work analyzing the security vulnerabilities and signal characteristics of Military Standard 1553 (MIL-STD-1553), the predominant military avionics bus. This research ranges from high-level overviews of potential attack vectors \cite{attack_vectors} to software-simulated buses \cite{sv1dur}. However, there is comparatively little research on ARINC 429 and AFDX, the primary standards used in commercial avionics, and recent work has more commonly focused on the more modern ethernet-based AFDX design\cite{asl}. \par
The sole study we were able to locate investigating the security of ARINC 429 focused on the feasibility of building an intrusion detection system (IDS). Gilboa-Markevich and Wool detailed a hardware fingerprinting method for ARINC 429 buses, utilizing changes in the electrical properties of the signal to detect intrusions \cite{ids}. This form of IDS typically uses machine learning-based anomaly detection algorithms, which have been demonstrated and tested on similar systems, such as Controller Area Network (CAN) buses \cite{can_ids}. An IDS would be capable of alerting the pilot to the type of attack presented in this paper, which focuses on ARINC 429's vulnerability to physical tampering. \par
The fidelity of the data on which a machine-learning IDS is trained and tested is an important consideration.  Inaccuracies and limitations resulting from sample bias will directly affect the system's ability to detect and classify rogue devices \cite{data_quality}. To optimize the real-world performance and reliability of systems that rely on machine learning, developers should aim to train and test them on large datasets of high quality, operationally relevant data with informative features. \par
Currently, there is a lack of publicly available ARINC message datasets, either from legitimate LRUs or from rogue devices. This represents a major barrier to the creation and refinement of IDSs, as anyone wishing to develop a high-performance avionics IDS would need to first construct an avionics test bed to gather data or collect data on-aircraft. Gilboa-Markevich and Wool trained and tested their IDS using ARINC messages generated by test and evaluation circuit boards \cite{ids}. These boards generate ARINC messages within specification, but are not commercial avionics equipment. Additionally, to prevent the IDS from unintentionally training on patterns in the message content, the ARINC messages they generated were designed to include all possible segment types. Finally, the messages are all even parity, and thus not valid according to the ARINC standard. The resulting dataset is not representative of the messages that actual avionics instrumentation on an ARINC 429 bus generates during normal operation. \par
We aimed to overcome the limitations of simulated data and the lack of operationally representative ARINC messages by creating a hardware-in-the loop (HITL) system for producing operationally relevant data. Specifically, we intended to achieve two goals with our HITL:
\begin{enumerate}
    \item Identify and verify ARINC 429 vulnerabilities
    \item Create publicly available ARINC 429 message datasets generated by avionics hardware and by hardware that could be used to conduct cyberattacks. 
\end{enumerate}

%\IEEEpubidadjcol

\subsection{Threat/Adversary Model}
As outlined in the previous section, the insecure design of ARINC 429 leaves it vulnerable to a large number of potential attacks. The wired nature of the ARINC bus generally requires that an adversary gain physical access to the system to conduct an attack. This should not be mistaken for a security feature, as an adversary can gain physical access to the bus on the aircraft (e.g., by posing as a technician) or via a supply-chain attack, adding malicious code or components to an LRU before it is installed in an aircraft. The increasing prevalence of Wi-Fi-connected devices on-aircraft increases the potential of an ARINC LRU being accessible remotely or via a passenger-facing component, such as the seat-back entertainment system \cite{digital_avionics}. Although Wi-Fi-connected devices do not affect the aircraft's safety, their ability to facilitate lateral movement between the aircraft's systems represents a clear security threat. \par
Attacks on an ARINC bus can be categorized by the type of LRU being targeted \cite{ids}:
\begin{itemize}
    \item \textbf{Rogue Transmitter}: In rogue transmitter attacks, the adversary targets the transmitter to send malicious messages to achieve cyber effects. The rogue transmitter lays dormant for a set period, imitating the behavior of the original transmitter to avoid detection. Once a given condition has been met the transmitter sends its malicious messages. The triggering condition can be passive, activating once a certain location, altitude, etc. has been reached, or active, via a radio frequency trigger. These messages could be invalid/random noise ("fuzzing"), previously recorded valid ARINC messages ("playback"), or custom-built, legitimate ARINC messages ("spoofing"). The effects of a rogue transmitter can range from relatively minor disruptions to potentially irreversible damage to electronic and physical components, including catastrophic loss of the aircraft.
    \item \textbf{Rogue Receiver}: In this scenario the adversary targets any of the receivers on the ARINC bus or adds their own receiver to the bus. While a rogue receiver attack is not as powerful or destructive as a rogue transmitter attack, an attacker could still exfiltrate important data. The rogue receiver could collect data containing critical flight or system information that would otherwise be unavailable. This information could then be used to facilitate further attacks across other channels. This is a clear violation of the Confidentiality leg of the Confidentiality, Integrity, Availability (CIA) triad \cite{cia}, as the adversary is an unauthorized individual accessing flight information.
\end{itemize}
An adversary cannot simply add their own transmitter to an ARINC bus, or convert a receiver to a transmitter, without violating the electrical properties of the system. The ARINC 429 bus is designed to allow exactly one transmitter LRU and no more than 20 receiver LRUs. Having multiple active transmitters on the bus will cause bus contention resulting in significant signal degradation in the best case and in permanent damage to the two bus drivers in the worst case. As a result, an adversary targeting a transmitter needs to replace the legitimate transmitter with their own. Alternatively, the adversary could insert a malicious circuit designed to disable the legitimate transmitter or to put it into a high impedance when the rogue transmitter is active on the bus, in order to prevent bus contention. Additionally, it is impossible to convert a receiver LRU to a transmitter purely through software manipulation, due to the bus standard. \par

\section{Methodology}
Our team created an HITL simulator with an ARINC 429 bus connecting a ground warning system to a multi-function display. An adversarial device was then connected to this bus, allowing us to replay messages to perform a denial-of-service attack against the navigational map of the multi-function display. We then used an oscilloscope to record voltage traces of both adversarial and valid ARINC 429 messages, creating a dataset for future use. \par
The attack demonstrated in this paper can either be viewed as a "technician" attack or as a supply-chain attack, as they are practically identical from a physical standpoint. In our attack, we connect a malicious transmitter to the bus via an electromechanical relay. There are other methods for implementing this switch; we chose an electromechanical relay because it was inexpensive and convenient. When the aircraft reaches a certain latitude (a "geo-fence" attack), the relay is triggered, disconnecting the legitimate transmitter from the bus and connecting the rogue transmitter to the bus.

\subsection{HITL Design}
\begin{figure}[!t]
\centering
\includegraphics[width=0.48\textwidth]{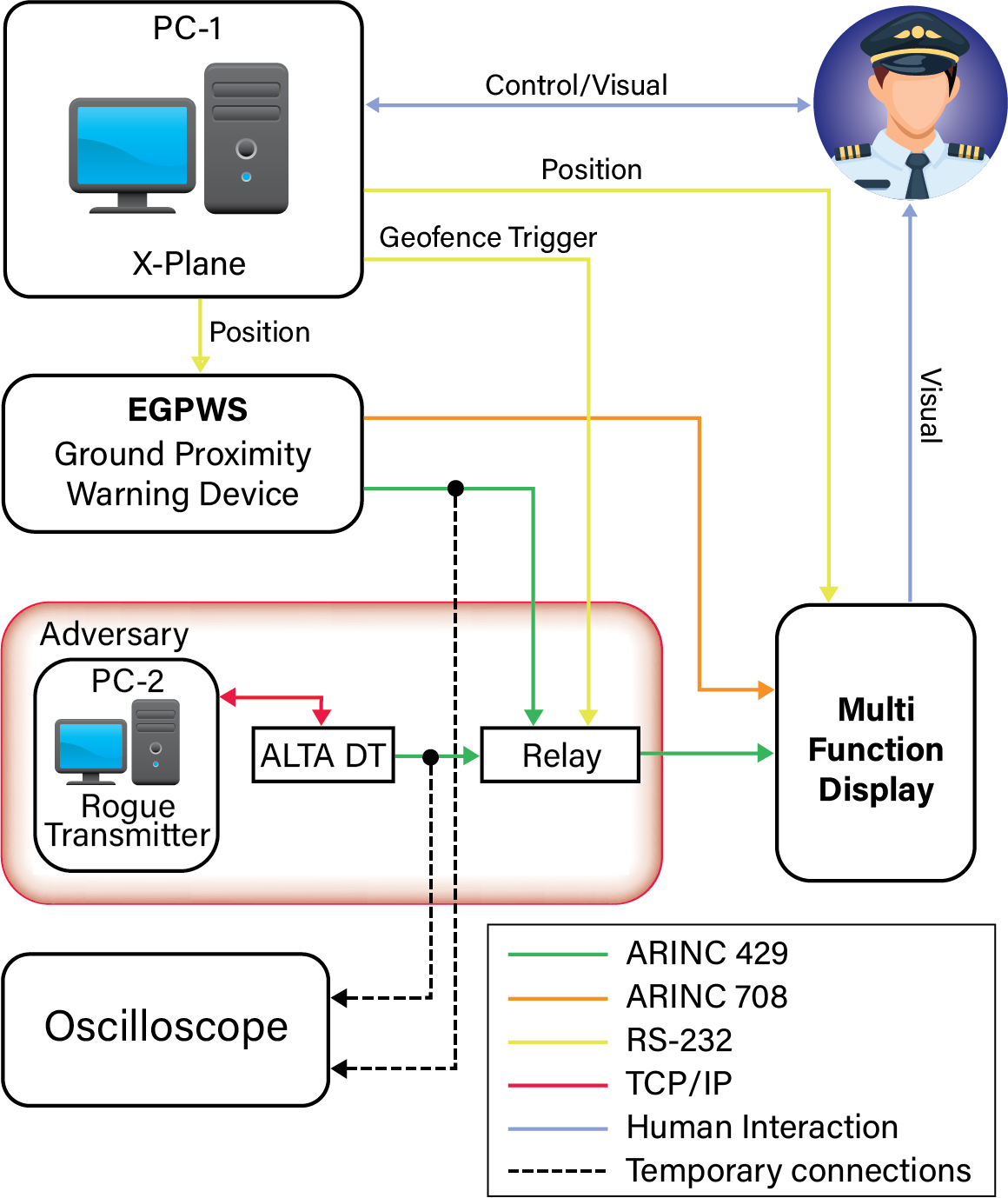}
\caption{Block diagram of HITL simulator}
\label{fig_1}
\end{figure}

Our HITL simulator is based on a Dell PC running X-Plane 11 Pro. A flight yoke, pedals, and throttle levers are used as its primary controls to enhance the realism of the flight experience. X-Plane exports flight data in real time to the Garmin GMX-200 Multi-Function Display (MFD) and the Honeywell KGP-500 Enhanced Ground Proximity Warning System (EGPWS) over RS-232 connections. In our simulator, these connections represent the data feeds from an aircraft’s instruments to the MFD and EGPWS. As shown in Fig. \ref{fig_1}, the EGPWS communicates to the pilot through the MFD, the cockpit’s speakers, and the annunciator. \par
The MFD is primarily used in our HITL for its moving map display, which enables the pilot to navigate and orient themselves in instrument flight rules (IFR) situations. The EGPWS communicates with the MFD via an ARINC 429 bus, where the EGPWS is the transmitter LRU and the MFD is a receiver LRU. The MFD is the target of our attack, which uses an electromechanical relay to replace the EGPWS transmitter LRU with signals from an AltaDT Enet-MA4 \cite{altadt}. The AltaDT Enet-MA4 is a small (13.5 x 3.7 x 4 cm), commercial-off-the-shelf diagnostic tool for MIL-STD-1553 and ARINC 429 buses. These qualities make the AltaDT tool representative of a leave-behind device that could be used in a cyberattack where the adversary has physical access to the ARINC 429 bus. It can send, receive, and store ARINC 429 messages, allowing it to function as either a receiver LRU or a transmitter LRU.  

\subsection{Data Collection}
Previous research on the security of ARINC 429 found that different transmitters exhibit different electrical properties, even if they are produced by the same manufacturer \cite{ids}. As a result, the voltage levels, or a function of the voltages (e.g. slope), of ARINC messages can be used to identify transmitters via machine learning algorithms. In addition to the voltage levels, we collected selected ARINC messages from actual avionics hardware to be used as messages in our replay attack. We used the AltaDT to record the messages transmitted across our HITL's bus while simultaneously recording the bus's voltage levels with a Textronix DPO 2002B oscilloscope. Using the oscilloscope, we were able to obtain precise voltage values for the messages generated by the EGPWS and the AltaDT box\footnote{The oscilloscope had a buffer size of 125,000 samples at 12.5MHz}. Only specific messages carrying terrain warnings needed to support the replay attack were recorded \footnote{The recorded messages were limited to those with label 270 and were Ground Proximity Warning System (GPWS) Discrete signals}. Voltage measurements were taken at the receive channels of the AltaDT to ensure that signal differences are caused only by differences between transmitters. 

\subsection{Attack Setup}
To conduct a playback attack, we first needed to record terrain warning messages sent by the EGPWS. To do this, we loaded X-Plane and proceeded to fly low to the ground, causing a terrain warning alert to appear on our MFD. X-Plane was then paused, which resulted in data from a fixed point in time being repeatedly transmitted. We recorded these terrain warning messages through the AltaDT tool before resuming the simulation and flying the aircraft up to 20,000 feet, so that the terrain warnings were no longer being issued. We then recorded a sample of EGPWS messages without terrain warnings. By comparing the two sets of messages and consulting the ARINC 429 specification, we determined that the terrain warning is controlled by a single bit in label 270 messages. This finding allowed us to trim down the set of messages that were being played back, increasing the rate at which terrain warning messages were sent. \par
We used the playback feature of the AltaDT box to constantly transmit a subset of our recorded EGPWS messages. This subset sends the terrain warning message every 100ms, ensuring that the MFD is saturated with terrain warning messages. However, the electromechanical relay prevents these messages from reaching the MFD until a certain trigger is met. In our attack, the relay is controlled by a geo-fence trigger within the Python code that exports X-plane data to our HITL over the serial RS-232 connections. In an attack scenario, this location data could be provided by a rogue receiver passively collecting GPS data from an ARINC 429 bus and triggering the rogue transmitter when a certain value is received.

\subsection{Attack Scenario}
The specific attack scenario simulated in this paper is a geo-fenced, playback denial-of-service attack with physical access to the ARINC 429 bus; however, the vulnerabilities demonstrated in this paper generalize to a much larger range of possible attacks. \par
Our scenario modeled an aircraft on approach to Dulles International Airport (KIAD) during severely overcast weather, resulting in an IFR situation. The simulation begins with the aircraft located at 76.65\degree W, 39.243\degree N, facing southwest. When the aircraft moves west of 77.3\degree W or south of 39.2\degree N, the serial port sends a signal to the electromechanical relay, causing it to disconnect the EGPWS from the ARINC 429 bus and instead connect the AltaDT box to the bus. Our simulated trigger is within the Python code communicating between X-plane and the HITL; however, in a real-world attack this trigger could come either from a dedicated GPS receiver built into the leave-behind device, or from a rogue receiver on the ARINC bus. 

\section{Results}
\subsection{Attack Effects}
One of the default configuration settings of the MFD is to automatically switch from the current display to the terrain map when a terrain warning message is received, as this should indicate imminent danger of a collision. When terrain warnings are maliciously looped at a high frequency, this default setting results in a denial-of-service attack against the pilot’s MFD, as the pilot is no longer able to access their moving map for navigational information. This attack does not affect the terrain information displayed on the terrain map, as that information is carried via a different ARINC bus designed for radar data (ARINC 708 \cite{arinc708}). \par
\begin{figure}[!t]
\centering
\includegraphics[width=0.4\textwidth]{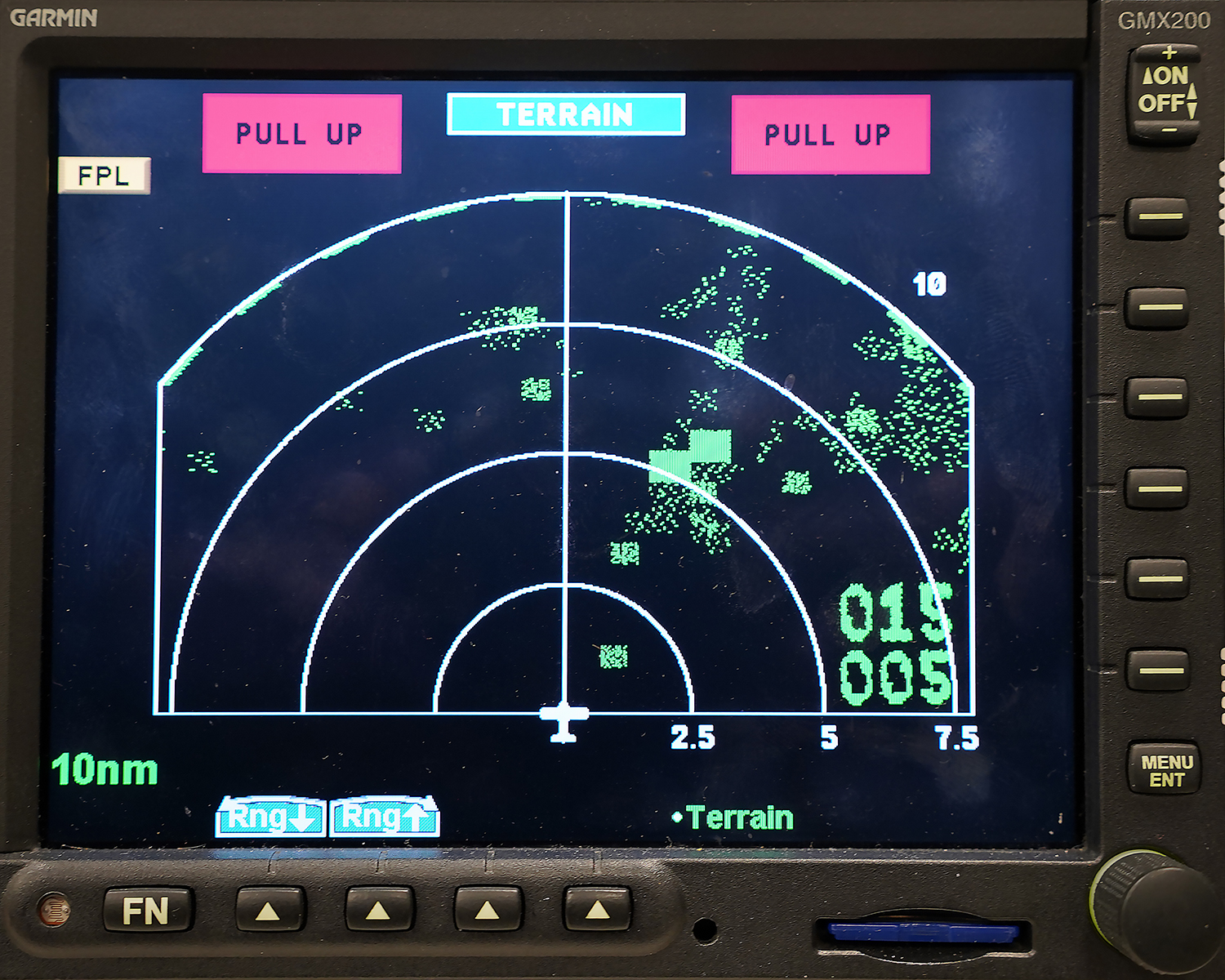}
\caption{Multi-Function Display showing terrain map with warnings during ARINC denial-of-service attack}
\label{fig_2}
\end{figure}

At this time, the only known method of defending against this attack in-flight is to restart the MFD twice; once to boot into the MFD’s configuration mode and disable the terrain awareness and warning system, and then again to load the new configuration. As shown in Fig. \ref{fig_2}, the MFD is displaying seemingly correct terrain information while incorrectly issuing constant terrain warnings, introducing further uncertainty into the pilot's decision making. This could cause the pilots to enter a state very similar to that of Mode Confusion, which occurs when an automated flight system behaves contrary to how the crew believes it should\cite{mode_confusion}. 

\subsection{Hardware-Generated ARINC 429 Data}
We collected two ARINC 429 derived datasets from our HITL while conducting this cyberattack: message logs and voltage traces. The message logs consist of hexadecimal representations of the binary messages being sent from the EGPWS and AltaDT to the MFD, along with timestamps indicating when these messages were sent. These hexadecimal messages can be decoded using the ARINC 429 standard (Appendix A), revealing insights into the data transmitted during normal operation. We identified a single bit in certain messages that is responsible for triggering terrain warning messages, which enabled us to refine our playback attack messages. \par
Creating a synthetic dataset of ARINC 429 messages can be accomplished using test and evaluation circuit boards, if one lacks the hardware to record actual ARINC 429 messages. Recreating the content of these messages in a simulated environment is trivial, as one can work backwards from the ARINC standard to create appropriate messages for the simulated bus. However, in reviewing these data from our hardware, we observed that our EGPWS was generating and transmitting more messages than the standard indicated it should be. It is not uncommon for avionics to violate the standards or the interface control documents used to describe the possible messages that could be sent on the bus. The presence of these additional messages reveals a critical shortcoming of purely simulated ARINC 429 buses. If an IDS were to be trained on messages from a simulated bus that only sent messages explicitly assigned to that piece of equipment in the standard, then when the IDS was applied to an actual bus it could trigger despite no intrusion being present. \par
\begin{figure}[!t]
\centering
\includegraphics[width=0.48\textwidth]{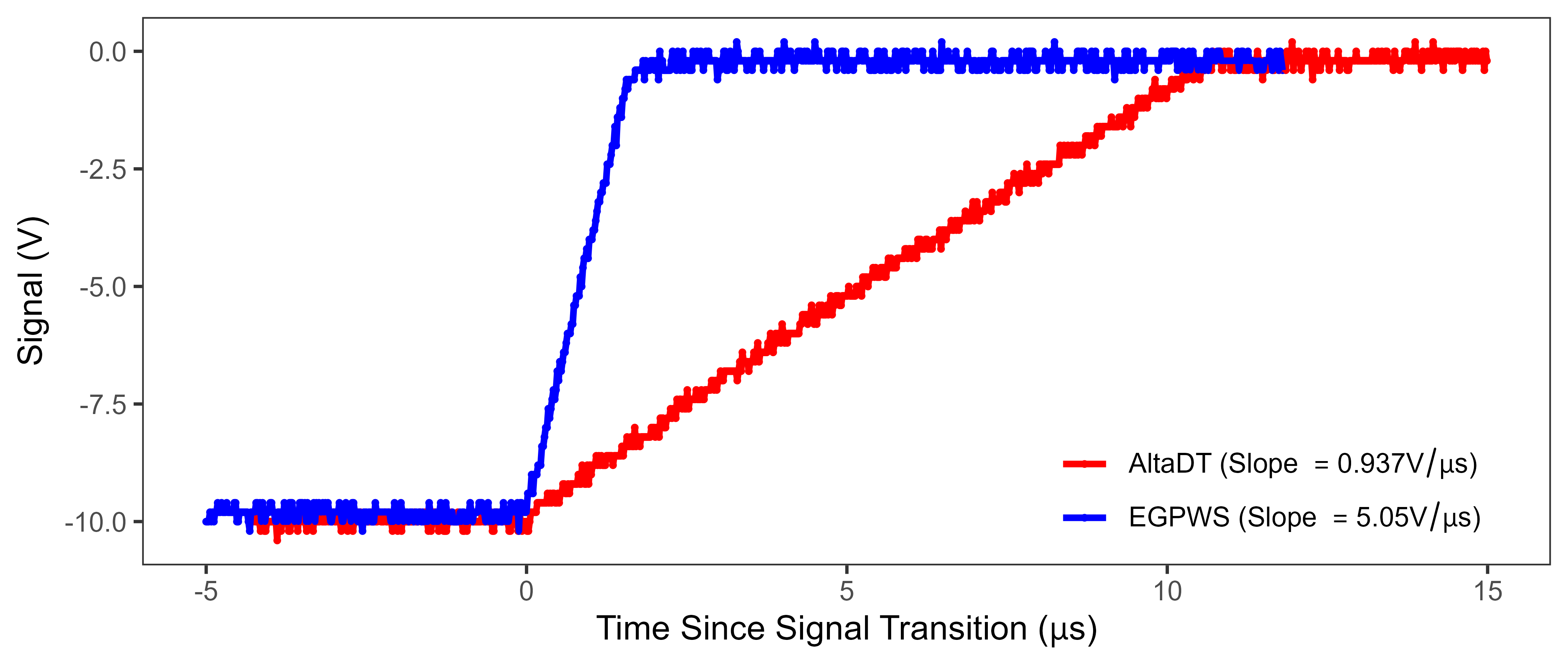}
\caption{EGPWS and AltaDT Time Response Comparison}
\label{fig_3}
\end{figure}

Voltage traces represent a significant difference in data fidelity between a simulated ARINC bus and a physical HITL ARINC bus, as they capture the electrical and timing characteristics of the signal. Although two different transmitters can send ARINC messages with identical content on a given bus, due to the protocol lacking any form of authentication, we could clearly observe differences in the electrical characteristics of messages sent by different transmitters. Messages sent by the EGPWS formed our baseline, and can be broadly characterized as having a quick transition between states (e.g. NULL to HI) with a slope of $5.05 V/\mu s$, as shown by the blue line in Fig. \ref{fig_3}. In contrast, messages sent by the AltaDT box (red line) had a much longer transition time, with a slope of $0.937 V/\mu s$. \par

The differences between the EGPWS and AltaDT messages are so pronounced that they can be visually observed on the oscilloscope, yet both are treated as valid messages by the receivers. \par
It is important to note that while our data demonstrates the signal differences between transmitters and is more operationally realistic than simulated data, we would not expect our results to exactly match the voltage traces from any given aircraft. The specific voltage characteristics of an ARINC 429 bus are unique to each configuration, as cable length and the number of receivers on a given bus will affect its impedance. This distinct fingerprint is how an IDS trained on the electrical properties of an ARINC 429 bus functions.

\section{Discussion}
\subsection{Broad Relevance of Study Results}
Due to the prevalence of ARINC 429 in commercial aviation, a compromised ARINC 429 bus could enable a wide variety of cyberattacks. Although our demonstrated scenario is limited, the underlying principles and methods behind the attack are widely generalizable:
\begin{itemize}
    \item \textbf{Penetration}: Our scenario features an adversary gaining physical access to the ARINC bus, whether that be with valid credentials or by avoiding detection from physical security. The adversary could also be an insider (someone with valid access to the aircraft, such as an airport employee) or be upstream in the supply chain. Loffi and Wallace \cite{insider_threat} identified airport employees as having nearly unlimited access to aircraft systems, as little emphasis is placed on monitoring these individuals.
    \item \textbf{Target}: Our attack targeted the ARINC 429 connection between the EGPWS and the MFD; however, attacks could target connections further upstream, such as between the GPS and EGPWS. Any systems connected via an ARINC 429 bus are vulnerable, provided that bus can be physically accessed. 
    \item \textbf{Attack Method}: Our scenario used a replay attack to transmit valid, prerecorded terrain warning messages during flight. Adversaries could also hand-craft their own ARINC messages, sending incorrect altitude/GPS data (spoofing), or send noise over the bus (fuzzing).
    \item \textbf{Attack Trigger}: A geo-fence attack was chosen in this scenario due to its ease of demonstration, because it is triggered without any direct input from the adversary. The range of possible attack triggers is incredibly vast, including both active and passive triggers.
    \item \textbf{Cyber Effects}: This scenario demonstrated a denial-of-service attack against the pilot's MFD, removing its functionality as a navigational aid. Other potential cyber effects of similar attacks include altering GPS or altitude data, which could result in operational effects such as the catastrophic loss of the aircraft.

\end{itemize} 

\subsection{ARINC 429 Vulnerability Mitigation}
With these vulnerabilities identified, we can explore potential solutions to mitigate the risks of a compromised ARINC 429 bus. One possible approach would be to add encryption or source authentication to the messages being sent, as these are commonplace in modern secure communication protocols. Despite the apparent simplicity of this approach, the implementation of these features is far from straightforward. Adding encryption at the software level would add timing delays to the signals, potentially compromising the safety of the system. These could be avoided by implementing encryption at the hardware level; however, this would require physically modifying every ARINC 429 bus currently in service. Given the widespread presence of ARINC 429 in commercial aircraft, replacing every bus currently on an active aircraft would be prohibitively costly and functionally infeasible. These issues make incorporating message authentication or encryption into the standard an unlikely mitigation option. Instead, solutions to ARINC 429 cyberattacks ideally work within the existing standard, minimizing the amount of time and resources needed for implementation.

\subsection{Importance of Hardware-Generated Data}
One way to mitigate cyberattacks on ARINC 429 buses is to install IDSs. IDSs can alert pilots or trigger automated actions to help secure the networks and restore CIA. IDSs have been implemented on similarly insecure buses, such as CAN buses \cite{can_ids}. \par
Machine learning can be used to design an effective IDS, assuming high quality training data are available in sufficient quantity. Gilboa-Markevich and Wool's work \cite{ids} demonstrates the effectiveness of a simulated ARINC 429 IDS; however, their IDS is trained and tested on message data created by test and evaluation integrated circuit boards, not avionics hardware.
The data they used were crafted to contain all possible message segment types within the ARINC 429 standard, while our data were obtained from a single piece of avionics. Because the leading edge of the positive pulse contains the necessary information to classify the source of the ARINC 429 traffic, and all messages will have positive, negative, and zero voltage segments, wide coverage across possible ARINC 429 message components would not be necessary for robustness. The datasets included with this publication are generated by actual avionics hardware using simulated flight data, representing an increase in data fidelity for future cybersecurity research.

\section{Conclusion}
Our research demonstrated a cyberattack on the ARINC 429 bus between an aircraft’s EGPWS and its MFD. This attack was performed within a custom HITL simulator, increasing its realism by demonstrating a denial-of-service attack on avionics hardware. While creating this attack, we produced multiple datasets of avionics-hardware-generated ARINC 429 data, which have been included with this publication to facilitate future study. \par
A goal for future work is to expand the number of avionics systems involved in our HITL, to enable us to generate insights into the effects of cyber compromise on other systems. This expansion would also enable investigation into the effects of compromising multiple systems simultaneously and of propagating attacks through ARINC 429 connections. For example, a compromised GPS could send incorrect data via ARINC 429 to its receiver LRUs, which include the EGPWS and MFD. The EGPWS would then unknowingly become a rogue transmitter to the MFD, as it would be performing terrain warning calculations using malicious data. Another avenue of future study is the testing of IDSs within the HITL and verifying their performance in an operationally representative setting. \par
ARINC 429 lacks any form of encryption or authentication, rendering it an insecure communication channel. Despite this, it is present on nearly every commercial fixed-wing aircraft built since the standard was introduced in 1977. Given its widespread reach, it is critical that the possibility of cyber compromise on such a system is not discounted, and thorough research should be performed to understand potential risks. Once these risks are understood, measures should be taken to develop and implement feasible solutions to mitigate its cyber vulnerabilities. Our research clearly demonstrated the danger of these vulnerabilities, and has generated a previously unavailable collection of realistic data for future work on detecting potential attacks.

\appendices
\section{ARINC Message Hexadecimal to Binary Conversion}
The ARINC 429 standard details the encoding for each message in binary format; however, AltaDT's Snapshot Viewer records ARINC 429 messages in hexadecimal format. As a result, translation between the two is a critical process, which we outline below with a worked example:
\subsection{Conversion Workflow}
\begin{enumerate}
    \item Convert hexadecimal message to binary
    \item Split binary string into component parts (Label, Source/Destination Identifier, Data, Sign/Status Matrix, Parity bit)
    \item Reverse the label field, so its bit order matches that of the rest of the message
    \item Convert label binary into octal
    \item Reference ARINC 429 standard for the data encoding given the label and equipment ID
    \item Apply specified encoding to data field, sign/status matrix
\end{enumerate}
\subsection{Worked Example}
\begin{enumerate}
    \item E189A8C1 = 1110 0001 1000 1001 1010 1000 1100 0001
    \item Label = 1100 0001, Source/Destination Identifier = 00, Data = 0 0001 1000 1001 1010 10, Sign/Status Matrix = 11, Parity bit = 1
    \item Label = 1000 0011
    \item Label = 10 000 011 = 203
    \item From \cite{arinc_std}, least significant data bit is padding, data is altitude (feet above mean sea level) in binary number representation format
    \item Data = 000 0110 0010 0110 101P, Data = 12,597 ft msl, Sign/Status Matrix: 11 = Normal Operation for BNR data
\end{enumerate}

\section{Voltage Trace File Format}
The voltage trace data is contained in two separate CSV files, one for messages received from the EGPWS and another for messages sent from the AltaDT box. The files formats are identical, consisting of the voltage traces from multiple messages, vertically stacked. The first row is the header, containing the names of each column. Descriptions of each column are listed below:
\begin{enumerate}
    \item {\textbf{[Index]}}: An integer value representing the position of each row in the dataframe, initialized at zero for the first non-header row
    \item \textbf{Time (s)}: A floating point value representing the time of each voltage sample with respect to the oscilloscope trigger point. Samples taken before the trigger have a value less than zero. The oscilloscope trigger is set for the rise of the first bit of the message, which occurs at approximately 2.5V.
    \item \textbf{Voltage (V)}: A floating point value representing the voltage differential between the two wires of the ARINC bus. This differential is calculated within the oscilloscope before the data is exported.
    \item \textbf{Word}: An integer representing which message each voltage sample comes from, initialized at one. Currently the EGPWS file contains 10 words and the AltaDT file contains 3.
\end{enumerate}

\section*{Acknowledgment}

The authors would like to thank Jason Schlup, Peter Mancini, Lee Allison, Evan Shockley, and Vincent Bass for their contributions to cybersecurity lab development at IDA.

\ifCLASSOPTIONcaptionsoff
  \newpage
\fi



\begin{thebibliography}{1}

\bibitem{arinc_std}
\emph{DIGITAL INFORMATION TRANSFER SYSTEM (DITS) PART 1 FUNCTIONAL DESCRIPTION, ELECTRICAL INTERFACES, LABEL ASSIGNMENTS, AND WORD FORMATS}, ARINC 429P1-19, SAE Industry Technologies Consortia, Bowie, MD, USA, 2019.

\bibitem{Fuchs2012TheEO}
Christian M. Fuchs and Stefan Schneele and Alexander Klein, "The Evolution of Avionics Networks From ARINC 429 to AFDX," 2012. [Online]. Available: {https://api.semanticscholar.org/CorpusID:14588951}

\bibitem{attack_vectors}
K. Lounis, Z. Mansour, M. Wrana, M. A. Elsayed, S. H. H. Ding and M. Zulkernine, "A Review and Analysis of Attack Vectors on MIL-STD-1553 Communication Bus," in \emph{IEEE Transactions on Aerospace and Electronic Systems}, vol. 58, no. 6, pp. 5586-5606, Dec. 2022, doi: {10.1109/TAES.2022.3177583}.

\bibitem{sv1dur}
J. Banks, R. Kerr, S. Ding and M. Zulkernine, "SV1DUR: A Real-Time MIL-STD-1553 Bus Simulator with Flight Subsystems for Cyber-Attack Modeling and Assessments," \emph{MILCOM 2022 - 2022 IEEE Military Communications Conference (MILCOM)}, Rockville, MD, USA, 2022, pp. 522-528, doi: {10.1109/MILCOM55135.2022.10017663}.

\bibitem{asl}
A. -V. Predescu and T. H. Stelkens-Kobsch, "Aviation Security Lab: A testbed for security testing of current and future aviation technologies," \emph{2022 IEEE/AIAA 41st Digital Avionics Systems Conference (DASC)}, Portsmouth, VA, USA, 2022, pp. 1-5, doi: {10.1109/DASC55683.2022.9925750}.

\bibitem{ids}
A. Wool and N. Gilboa-Markevich, "Hardware Fingerprinting for the ARINC 429 Avionic Bus," in \emph{ESORICS 2020. Lecture Notes in Computer Science(), vol 12309.}, L. Chen, N. Li, K. Liang and S. Schneider, Ed., Springer, Cham. {https://doi.org/10.1007/978-3-030-59013-0\_3}

\bibitem{digital_avionics}
K. Kainrath, M. Fruhmann, K. Gebeshuber, E. Leitgeb and M. Gruber, "Evaluation of Cyber Security in Digital Avionic Systems," \emph{2020 IEEE 91st Vehicular Technology Conference (VTC2020-Spring)}, Antwerp, Belgium, 2020, pp. 1-5, doi: {10.1109/VTC2020-Spring48590.2020.9128447}.

\bibitem{mode_confusion}
A. Joshi, S. P. Miller and M. P. E. Heimdahl, "Mode confusion analysis of a flight guidance system using formal methods," \emph{Digital Avionics Systems Conference, 2003. DASC '03. The 22nd}, Indianapolis, IN, USA, 2003, pp. 2.D.1-21-12 vol.1, doi: {10.1109/DASC.2003.1245813}.

\bibitem{insider_threat}
J.M. Loffi and R.J. Wallace, "The unmitigated insider threat to aviation (Part 1): a qualitative analysis of risks," J Transp Secur 7, pp. 289–305 [2014]. https://doi.org/10.1007/s12198-014-0144-4

\bibitem{can}
\emph{CAN Specification 2.0}, ROBERT BOSCH GmbH, Postfach 50, D-7000 Stuttgart 1, Germany, 1991. [Online]. Available: http://esd.cs.ucr.edu/webres/can20.pdf

\bibitem{uart}
\emph{SCC2691: Universal asynchronous receiver/transmitter (UART)}, Phillips Semiconductors, 2006. [Online]. Available: https://www.nxp.com/docs/en/data-sheet/SCC2691.pdf

\bibitem{altadt}
\emph{eNet-MA4}, Alta Data Technologies, 4901 Rockaway Blvd., Rio Rancho, NM, USA 2022. [Online]. Available: https://www.altadt.com/product/enet-ma4-1553-arinc-ethernet-converter/

\bibitem{arinc708}
\emph{708-6 Airborne Weather Radar (WXR)}, ARINC 708-6, SAE Industry Technologies Consortia, Bowie, MD, USA, 1991. 

\bibitem{can_ids}
W. Choi, K. Joo, H. J. Jo, M. C. Park and D. H. Lee, "VoltageIDS: Low-Level Communication Characteristics for Automotive Intrusion Detection System," in \emph{IEEE Transactions on Information Forensics and Security}, vol. 13, no. 8, pp. 2114-2129, Aug. 2018, doi: 10.1109/TIFS.2018.2812149.

\bibitem{afdx}
\emph{AIRCRAFT DATA NETWORK PART 7 AVIONICS FULL DUPLEX SWITCHED ETHERNERT (AFDX) NETWORK}, ARINC 664P7, SAE Industry Technologies Consortia, Bowie, MD, USA, 2005. 

\bibitem{data_quality}
S.E. Whang, Y. Roh, H. Song, et al. "Data collection and quality challenges in deep learning: a data-centric AI perspective." \emph{The VLDB Journal} 32, 791–813 (2023). https://doi.org/10.1007/s00778-022-00775-9

\bibitem{cia}
S. Samonas and D. Coss, "The CIA strikes back: Redefining confidentiality, integrity and availability in security." \emph{Journal of Information System Security}, vol. 10, issue 3, 2014. [Online]. Available: https://www.proso.com/dl/Samonas.pdf

\end{thebibliography}
\end{document}